\documentstyle[12pt,fleqn]{article}

\def\thebibliography#1{\section*{References}\list
  {[\arabic{enumi}]}{\settowidth\labelwidth{#1}\leftmargin\labelwidth
    \advance\leftmargin\labelsep
    \usecounter{enumi}}
    \def\newblock{\hskip .11em plus .33em minus .07em}
    \sloppy\clubpenalty4000\widowpenalty4000
    \sfcode`\.=1000\relax}

\def\op#1{\mathop{\fam0 #1}\limits}

\newcommand{\id}{{\rm Id\,}}
\newcommand{\Ker}{{\rm Ker\,}}
\newcommand{\der}{{\rm Der\,}}
\newcommand{\im}{{\rm Im\,}}
\newcommand{\nm}[1]{\mid {#1}\mid}
\newcommand{\nw}[1]{[{#1}]}
\newcommand{\pr}{{\rm pr}}

\newcommand{\beq}{\begin{equation}}
\newcommand{\eeq}{\end{equation}}
\newcommand{\ben}{\begin{eqnarray}}
\newcommand{\een}{\end{eqnarray}}
\newcommand{\be}{\begin{eqnarray*}}
\newcommand{\ee}{\end{eqnarray*}}
\newcommand{\bea}{\begin{eqalph}}
\newcommand{\eea}{\end{eqalph}}

\newcommand{\cA}{{\cal A}}

\newcommand{\cP}{{\cal P}}
\newcommand{\cR}{{\cal R}}
\newcommand{\cL}{{\cal L}}
\newcommand{\cE}{{\cal E}}
\newcommand{\cH}{{\cal H}}
\newcommand{\cF}{{\cal F}}
\newcommand{\cN}{{\cal N}}
\newcommand{\cV}{{\cal V}}
\newcommand{\cS}{{\cal S}}
\newcommand{\bL}{{\bf L}}
\newcommand{\bQ}{{\bf Q}}
\newcommand{\bC}{{\bf C}}
\newcommand{\bR}{{\bf R}}
\newcommand{\bom}{{\bf\Omega}}
\newcommand{\bth}{{\bf\Theta}}
\newcommand{\al}{\alpha}

\newcommand{\dl}{\delta}
\newcommand{\up}{\upsilon}
\newcommand{\la}{\lambda}
\newcommand{\La}{\Lambda}
\newcommand{\f}{\phi}
\newcommand{\om}{\omega}
\newcommand{\Om}{\Omega}
\newcommand{\m}{\mu}

\newcommand{\z}{\zeta}
\newcommand{\g}{\gamma}
\newcommand{\G}{\Gamma}
\newcommand{\th}{\theta}

\newcommand{\vt}{\vartheta}
\newcommand{\si}{\sigma}
\newcommand{\w}{\wedge}
\newcommand{\wt}{\widetilde}
\newcommand{\wh}{\widehat}
\newcommand{\ol}{\overline}
\newcommand{\dr}{\partial}
\newcommand{\ar}{\op\longrightarrow}
\newcommand{\ot}{\otimes}

\newcounter{eqalph}
\newcounter{equationa}
\newcounter{theorem}
\newcounter{remark}
\newcounter{proposition}
\newcounter{lemma}
\newcounter{corollary}
\newcounter{definition}

\newenvironment{eqalph}{\stepcounter{equation}
\setcounter{equationa}{\value{equation}}
\setcounter{equation}{0}

\begin{eqnarray}}{\end{eqnarray}\setcounter{equation}{\value{equationa}}}

\def\theremark{\arabic{remark}}

\def\thedefinition{\arabic{definition}}

\newenvironment{proof}{\noindent 
{\it Proof.}}{\medskip}
\newenvironment{rem}{\refstepcounter{remark}\medskip\noindent{\it
Remark \theremark.}}{\medskip}
\newenvironment{theo}{\refstepcounter{definition} 
\bigskip\noindent{\it Theorem \thedefinition.}}{\medskip}
\newenvironment{prop}{\refstepcounter{definition} 
\bigskip\noindent{\it Proposition \thedefinition.}}{\medskip}
\newenvironment{lem}{\refstepcounter{definition} 
\bigskip\noindent{\it Lemma \thedefinition.}}{\medskip}
\newenvironment{cor}{\refstepcounter{definition} 
\bigskip\noindent{\it Corollary \thedefinition.}}{\medskip}

\newcommand{\mar}[1]{}

\begin{document}
\hbox{}

{\parindent=0pt

{\large\bf Constraints in polysymplectic (covariant) Hamiltonian formalism}
\bigskip

{\sc G. Sardanashvily}
\medskip

Department of Theoretical Physics, Physics Faculty, Moscow State
University, 117234 Moscow, Russia; E-mail: sard@grav.phys.msu.su
\bigskip

\begin{small}
{\bf Abstract.}
In the framework of polysymplectic Hamiltonian formalism,
degenerate Lagrangian field systems are described as multi-Hamiltonian
systems with Lagrangian constraints. The physically relevant case of
degenerate quadratic Lagrangians is analyzed in detail, and the
Koszul--Tate resolution of Lagrangian constraints is constructed in an
explicit form. The particular case of Hamiltonian mechanics with
time-dependent constraints is studied.

\end{small}
}

\section{Introduction}

Let $Y\to X$ be a smooth fibre bundle of a classical field theory. We
consider first order Lagrangian field systems whose configuration
space is the first order jet manifold $J^1Y$ of sections of $Y\to X$.  
Polysymplectic Hamiltonian formalism enables us to describe
these systems as constraint Hamiltonian
systems on the Legendre bundle
\mar{00}\beq
\Pi=\op\w^nT^*X\op\ot_YV^*Y\op\ot_YTX \label{00}
\eeq
\cite{book,jpa,sard94,sard95}. Given fibred coordinates $(x^\la,y^i)$ on $Y$,
the Legendre bundle $\Pi$ is provided with the holonomic coordinates
$(x^\la,y^i,p^\la_i)$. Every Lagrangian 
\mar{cmp1}\beq
L=\cL\om: J^1Y\to\op\w^nT^*X, \quad \om=dx^1\w\cdots \w dx^n, \quad n=\dim X,
\label{cmp1}
\eeq
on $J^1Y$ yields the Legendre map
\mar{t1}\beq
\wh L:J^1Y \op\to_Y \Pi, \qquad p^\la_i\circ\wh L
=\pi^\la_i=\dr^\la_i\cL. \label{t1} 
\eeq
Lagrangian formalism on $J^1Y$ and polysymplectic
Hamiltonian formalism on $\Pi$ are equivalent if a Lagrangian $L$ is
hyperregular, i.e., $\wh L$ is a diffeomorphism.
In Part I of the work, we study the case of almost regular Lagrangians
$L$ when: 
(i) the Lagrangian constraint space $N_L=\wh L(J^1Y)$  is a closed
imbedded subbundle $i_N:N_L\hookrightarrow\Pi$ of the Legendre bundle $\Pi\to
Y$ and (ii) the Legendre map 
\mar{cmp12}\beq
\wh L:J^1Y\to N_L \label{cmp12}
\eeq
is a fibred manifold with connected fibres.
Lagrangians of the most of
field models are of this type. From the mathematical viewpoint, this notion of
degeneracy is particulary appropriate in order to study Lagrangian
constraints in  
polysymplectic Hamiltonian formalism. 
In this case, there are comprehensive relations
between Euler--Lagrange and Cartan equations in Lagrangian formalism,
Hamilton--De Donder 
equations in multisymplectic Hamiltonian formalism, covariant Hamilton
equations and  constrained Hamilton equations in polysymplectic
Hamiltonian formalism (see Theorems \ref{3.23}, \ref{3.24} and \ref{3.02}
below). The main peculiarity of these
relations lies in the fact that a
set of Hamiltonian 
forms is associated to a degenerate Lagrangian.

In Part II, we provide the detailed analysis  
of systems with degenerate quadratic Lagrangians, appropriate
for application to many physical models. Such a Lagrangian $L$ yields 
splittings of the affine jet bundle $J^1Y\to Y$ and the Legendre bundle
$\Pi\to Y$ (see Theorem \ref{ten2} below). 
The corresponding projection operators enable us to construct the
Koszul--Tate resolution of the Lagrangian constraints $N_L$ in an
explicit form. 

If $X=\bR$, polysymplectic Hamiltonian formalism provides the adequate
formulation of Hamiltonian time-dependent mechanics
\cite{book98,sard98}. Part III of the work is devoted 
to mechanical systems with time-dependent constraints.
The key point lies in the fact that, in time-dependent mechanics, the
canonical Poisson structure does not provide dynamic equations and the
Poisson bracket of constraints with a Hamiltonian is ill-defined  \cite{jmp00}.
\bigskip\bigskip

\begin{center}

{\large \bf PART I. Lagrangian constraints}

\end{center}
\bigskip\bigskip

All maps throughout are smooth, while manifolds are real,
finite-dimensional, Hausdorff, second-countable and connected. The 
$s$-order jet manifold $J^sY$ of a fibre bundle $Y\to X$ is endowed
with the adapted  
coordinates $(x^\la,y^i_\La)$, $0\leq\mid\La\mid\leq s$, where 
$\La$ is a multi-index $(\la_k...\la_1)$, $\nm\La=k$. 
We denote by
\be
h_0:\f_\la dx^\la+ \f_i^\La dy^i_\La \mapsto (\f_\la
+\f_i^\La y^i_{\la+\La})dx^\la
\ee
the exterior algebra
homomorphism which sends exterior forms on $J^sY$ onto 
horizontal forms on
$J^{s+1}Y\to X$
and vanishes on contact forms $\th^i_\La=dy^i_\La
-y^i_{\la+\La}dx^\la$. Let 
$d_\la =
\dr_\la +y^i_{\la+\La}\dr^\La_i$ be the  total derivative 
and 
$d_H\f=dx^\la\w d_\la\f$ the horizontal differential such that
$h_0\circ d=d_H\circ h_0$. A connection on a fibre bundle $Y\to X$ is
regarded as a
global section 
\be
\G=dx^\la\ot(\dr_\la +\G^i_\la\dr_i) 
\ee
of the affine jet bundle $\pi^1_0:J^1Y\to Y$.
Sections of the underlying vector bundle $T^*X\op\ot_YVY\to Y$ are called
soldering forms. 

\section{Lagrangian dynamics}

This Section and Section 3
summarize the basic notions of Lagrangian and polysymplectic
Hamiltonian formalisms 
(see \cite{book,jpa,sard95} for a detailed exposition).

Given a Lagrangian $L$ and its Lepagean
equivalent
$H_L$, the first variational formula of the
calculus of variations provides the canonical decomposition of the Lie
derivative 
of $L$ along a projectable vector field $u$ on $Y$:
\mar{C30}\beq
\bL_{J^1u}L=
 u_V\rfloor \cE_L + d_Hh_0(u\rfloor H_L), \label{C30} 
\eeq
where $u_V=(u\rfloor\th^i)\dr_i$ and
\mar{305}\beq
\cE_L=
 (\dr_i- d_\la\dr^\la_i)\cL \th^i\w\om: J^2Y\to T^*Y\w(\op\w^nT^*X) \label{305}
\eeq
is the Euler--Lagrange operator.
The kernel of $\cE_L$ is the Euler--Lagrange equations 
\mar{b327}\beq
(\dr_i- d_\la\dr^\la_i)\cL=0. \label{b327} 
\eeq

We will restrict our consideration to
the Poincar\'e--Cartan form
\mar{303}\beq
 H_L=L +\pi^\la_i\th^i\w\om_\la, \qquad
\om_\la=\dr_\la\rfloor\om. \label{303}
\eeq
It is a Lagrangian counterpart of Hamiltonian forms in polysymplectic
Hamiltonian formalism. Being a Lepagean equivalent of the Lagrangian
$L=h_0(H_L)$ on $J^1Y$, this is 
also a Lepagean equivalent of the Lagrangian 
\mar{cmp80}\beq
\ol L=\wh h_0(H_L) = (\cL + (\wh y_\la^i - y_\la^i)\pi_i^\la)\om, \qquad \wh
h_0(dy^i)=\wh y^i_\la dx^\la, 
\label{cmp80}
\eeq
on the repeated jet manifold $J^1J^1Y$ coordinated 
by $(x^\la,y^i,y^i_\la,\wh y^i_\la,y^i_{\la\m})$.  The
Euler--Lagrange operator for $\ol L$ reads
\mar{2237}\ben
&& \cE_{\ol L} : J^1J^1Y\to T^*J^1Y\w(\op\w^n T^*X), \nonumber \\
&& \cE_{\ol L} = [(\dr_i\cL - \wh d_\la\pi_i^\la 
+ \dr_i\pi_j^\la(\wh y_\la^j - y_\la^j))dy^i + \dr_i^\la\pi_j^\m(\wh
y_\m^j - y_\m^j) dy_\la^i]\w \om, \label{2237} 
\een
where $\wh d_\la=\dr_\la +\wh y^i_\la\dr_i +y^i_{\la\m}\dr_i^\m$.
Its kernel
$\Ker\cE_{\ol L}\subset J^1J^1Y$ is the Cartan equations
\mar{b336}\beq
\dr_i^\la\pi_j^\m(\wh y_\m^j - y_\m^j)=0, \qquad
\dr_i \cL - \wh d_\la\pi_i^\la
+ (\wh y_\la^j - y_\la^j)\dr_i\pi_j^\la=0. \label{b336}
\eeq
On sections $\ol s: X\to J^1Y$, the Cartan equations (\ref{b336})
are equivalent to the condition
\mar{C28}\beq
 \ol s^*(u\rfloor dH_L)=0 \label{C28}
\eeq
for any vertical vector field $u$ on $J^1Y\to X$. The Cartan equations
are equivalent to the Euler--Lagrange equations on integrable sections
$\ol s=J^1s$ of $J^1Y\to X$. 

The Poincar\'e--Cartan form $H_L$ (\ref{303}) yields the  
Legendre morphism
\be
\wh H_L: J^1Y\op\to_Y Z_Y, \qquad 
(p^\m_i, p)\circ\wh H_L =(\pi^\m_i, \cL-\pi^\m_i y^i_\m ), 
\ee
of
$J^1Y$ to the homogeneous Legendre bundle
\mar{N41}\beq
Z_Y= T^*Y\w(\op\w^{n-1}T^*X)  \label{N41}
\eeq
equipped with the holonomic
coordinates $(x^\la,y^i,p^\la_i,p)$.
There is the 1-dimensional affine bundle
\mar{b418'}\beq
\pi_{Z\Pi}:Z_Y\to \Pi \label{b418'}
\eeq
modelled over the pull-back vector bundle
$\Pi\op\times_X\op\w^nT^*X\to \Pi$. We have
\mar{ten3}\beq
\wh L=\pi_{Z\Pi}\circ \wh H_L. \label{ten3}
\eeq
Due to the monomorphism $Z_Y\hookrightarrow \op\w^nT^*Y$, the bundle
$Z_Y$ is endowed with the pull-back
\beq
\Xi_Y= p\om + p^\la_i dy^i\w\om_\la \label{N43}
\eeq
of the canonical form $\Theta$ on 
$\op\w^nT^*Y$ whose exterior differential $d\Theta$ is the $n$-multisymplectic
form in the sense of Martin \cite{mrt}. 

Let $Z_L=\wh H_L(J^1Y)$  be an imbedded subbundle
$i_L:Z_L\hookrightarrow Z_Y$ of $Z_Y\to Y$.
It is provided with the pull-back 
De Donder form $\Xi_L=i^*_L\Xi_Y$. We have 
\mar{cmp14}\beq
H_L=\wh H_L^*\Xi_L=\wh H_L^*(i_L^*\Xi_Y).  \label{cmp14}
\eeq
By analogy with the Cartan equations
(\ref{C28}), the  Hamilton--De Donder equations for sections $\ol r$ of 
$Z_L\to X$ are written as
\mar{N46}\beq
\ol r^*(u\rfloor d\Xi_L)=0 \label{N46}
\eeq
where $u$ is an arbitrary vertical vector field on
$Z_L\to X$. 

\begin{theo}\label{ddd} \mar{ddd}  Let the Legendre morphism
$\wh H_L$ be a submersion. Then a section $\ol s$ of $J^1Y\to X$
is a solution of the Cartan equations (\ref{C28}) iff $\wh
H_L\circ\ol s$ is a solution of the Hamilton--De Donder equations
(\ref{N46}) \cite{got91}.
\end{theo}

\section{Polysymplectic Hamiltonian dynamics}

The canonical polysylplectic form $\Om$, Hamiltonian connections and
Hamiltonian 
forms are the main ingredients in the covariant Hamiltonian dynamics on
the Legendre bundle
\be
\pi_{\Pi X}=\pi\circ\pi_{\Pi Y}:\Pi\to Y\to X.
\ee

Let us consider the canonical bundle monomorphism 
\mar{2.4}\beq
\th =-p^\la_idy^i\w\om\otimes\dr_\la
:\Pi\op\hookrightarrow_Y\op\w^{n+1}T^*Y\op\otimes_Y TX. \label{2.4}
\eeq
The polysymplectic form on
$\Pi$ is defined as a unique
$TX$-valued $(n+2)$-form
\mar{406}\beq
\Om =dp_i^\la\w dy^i\w \om\ot\dr_\la \label{406}
\eeq
such that the relation
$\Om\rfloor\f =-d(\th\rfloor\f)$
holds for any exterior 1-form $\f$ on $X$. 
A connection 
\be
\g =dx^\la\otimes(\dr_\la +\g^i_\la\dr_i
+\g^\m_{\la i}\dr^i_\m) 
\ee
on $\Pi\to X$ is called a Hamiltonian
connection if the exterior form $\g\rfloor\Om$
is closed. 
A Hamiltonian form $H$ on $\Pi$ is defined as the pull-back 
\mar{b418}\beq
 H=h^*\Xi_Y= p^\la_i dy^i\w \om_\la -\cH\om \label{b418}
\eeq
of the canonical form $\Xi_Y$ (\ref{N43}) by a 
section $h$ of the affine bundle (\ref{b418'}). 
Hamiltonian forms on $\Pi$ constitute an affine space modelled
over the linear space of horizontal densities $\wt H=\wt{\cH}\om$
on $\Pi\to X$. 

\begin{theo} \cite{book}.
For every Hamiltonian form $H$ (\ref{b418}), there exists an associated 
Hamiltonian connection 
such that
\mar{cmp3}\beq
\g\rfloor\Om= dH, \qquad
\g^i_\la=\dr^i_\la\cH, \qquad \g^\la_{\la i}= -\dr_i\cH. \label{cmp3}
\eeq
Conversely, for any Hamiltonian connection 
$\g$, there exists a local Hamiltonian form
$H$ on a neighbourhood of any point $q\in\Pi$ such that the relations
(\ref{cmp3}) hold. 
\end{theo}

For instance, every connection
$\G$ on $Y\to X$ defines the section
\be
h_\G: \ol dy^i \mapsto dy^i- \G^i_\la dx^\la
\ee
of the affine bundle $Z_Y\to\Pi$ and the corresponding
Hamiltonian form 
\mar{3.6}\ben
H_\G =h_\G^*\Xi_Y =p^\la_i dy^i\w\om_\la -p^\la_i\G^i_\la\om. \label{3.6}
\een
As a consequence, every 
Hamiltonian form $H$ admits the decomposition
\mar{4.7}\beq
H=H_\G -\wt H_\G =p^\la_idy^i\w\om_\la
-p^\la_i\G^i_\la\om-\wt{\cH}_\G\om. \label{4.7}
\eeq

Any bundle morphism
\mar{2.7}\beq
\Phi=dx^\la\otimes(\dr_\la +\Phi^i_\la\dr_i):\Pi\op\to_Y J^1Y,
\label{2.7}
\eeq
called a Hamiltonian map, defines the Hamiltonian form
\mar{jpa3}\beq
H_\Phi=\Phi\rfloor\th =p^\la_idy^i\w\om_\la -p^\la_i\Phi^i_\la\om.
\label{jpa3}
\eeq
Every Hamiltonian form $H$ (\ref{b418}) yields the Hamiltonian
map $\wh H$ such that
$y_\la^i\circ\wh H=\dr^i_\la\cH$. A Hamiltonian form $H$ is called
degenerate if the Hamiltonian map $\wh H$ is degenerate.

A Hamiltonian form $H$ (\ref{b418}) on $\Pi$ can be
seen as the Poincar\'e--Cartan form of the  Lagrangian
\mar{Q3}\beq
L_H = (p^\la_iy^i_\la - \cH)\om \label{Q3}
\eeq
on the jet manifold $J^1\Pi$. The Euler--Lagrange operator
(\ref{305}) for $L_H$, called the Hamilton operator for $H$, is  
\mar{3.9}\ben
&& \cE_H :J^1\Pi\to T^*\Pi\w(\op\w^n T^*X),\nonumber \\
&& \cE_H=[(y^i_\la-\dr^i_\la\cH) dp^\la_i
-(p^\la_{\la i}+\dr_i\cH) dy^i]\w\om. \label{3.9}
\een 
Its kernel is the covariant Hamilton equations
\mar{b4100}\bea
&& y^i_\la=\dr^i_\la\cH, \label{b4100a}\\
&& p^\la_{\la i}=-\dr_i\cH.\label{b4100b}
\eea
It is readily observed that all Hamiltonian connections (\ref{cmp3})
associated with a Hamiltonian form $H$ live in the kernel of the Hamilton
operator
$\cE_H$. Consequently, every integral section 
$J^1r=\g\circ r$
of a Hamiltonian connection $\g$ associated with a Hamiltonian form $H$ is
a solution of the Hamilton equations (\ref{b4100a}) --
(\ref{b4100b}). 
Similarly to the Cartan equations (\ref{C28}), the Hamilton
equations (\ref{b4100a}) -- (\ref{b4100b})  are equivalent to the condition
\mar{N7}\beq
r^*(u\rfloor dH)= 0 \label{N7}
\eeq
for any vertical vector field $u$ on $\Pi\to X$.

\begin{rem}
Lagrangians (\ref{Q3}) play an important role in the path integral
quantization of covariant Hamiltonian field theories \cite{ijtp}
\end{rem}

\section{Degenerate systems}

Let us state the relations between 
Lagrangian and polysymplectic Hamiltonian formalisms when a Lagrangian is 
degenerate (see \cite{book,jpa} for a detailed exposition). 

Given a Lagrangian $L$, a Hamiltonian form $H$ is said to be associated
with $L$ if the relations
\mar{2.30}\bea
&&\wh L\circ\wh H\circ \wh L=\wh L,\label{2.30a} \\
&&H=H_{\wh H}+\wh H^*L \label{2.30b}
\eea
hold.
A glance at the relation (\ref{2.30a}) shows that $\wh
L\circ\wh H$ is a projection
\beq
p^\m_i(q)=\dr^\m_i\cL (x^\m,y^i,\dr^j_\la\cH(q)), \qquad q\in N_L,
\label{b481'}
\eeq
of $\Pi$ onto the Lagrangian constraint space $N_L=\wh L( J^1Y)$.
Accordingly,  $\wh H\circ\wh L$ is a projection of $J^1Y$ onto $\wh
H(N_L)$. A Hamiltonian form is called weakly associated with a Lagrangian $L$
if the condition (\ref{2.30b}) holds on the Lagrangian constraint space $N_L$.

Given a Lagrangian $L$, one can construct associated and weakly
associated Hamiltonian forms as follows.

\begin{prop} \label{jp} \mar{jp} \cite{book}.
If a Hamiltonian map $\Phi$ (\ref{2.7}) obeys the relation (\ref{2.30a}),
then the Hamiltonian form $H=H_\phi+\Phi^*L$ is weakly associated with the
Lagrangian $L$. If $\Phi=\wh H$, then $H$ is associated with $L$.
\end{prop}

Hamiltonian forms weakly associated with a Lagrangian $L$ have the
following common property \cite{book}.

\begin{prop} \label{cmp110} \mar{cmp110}
Restricted to the Lagrangian constraint space $N_L$, any Hamiltonian form $H$ 
weakly associated with a Lagrangian $L$ coincides with the pull-back
\be
H\mid_{N_L}=\wh H^*H_L\mid_{N_L}
\ee
of the Poincar\'e--Cartan form $H_L$ (\ref{303}) by the
Hamiltonian map $\wh H$.
\end{prop}

Note that the essential difference between associated and weakly
associated Hamiltonian forms lies in the fact that, as follows from the
relation (\ref{2.30b}), associated Hamiltonian forms are necessarily
degenerate outside a Lagrangian constraint space.
Further, we study weakly associated Hamiltonian forms.

Let us restrict our consideration to almost regular Lagrangians.
In this case, Proposition \ref{jp} leads to the following criterion of
the existence of weakly associated Hamiltonian forms.

\begin{prop} \label{ten4} \mar{ten4}
A Hamiltonian form weakly associated with an almost regular Lagrangian
$L$ exists iff the fibred manifold $J^1Y\to N_L$ admits a global section.
\end{prop}

The following property of almost regular Lagrangians plays an important
role in the sequel \cite{book}.

\begin{lem} \label{3.22} \mar{3.22}
The Poincar\'e--Cartan form $H_L$ for an almost regular Lagrangian $L$
is constant on connected fibres of the fibred manifold $J^1Y\to N_L$.
\end{lem}

Then we come to the following assertion \cite{book}.

\begin{prop} \label{3.22'} \mar{3.22'} All Hamiltonian forms weakly associated
with an almost regular Lagrangian $L$ coincide with each
other on the Lagrangian constraint space $N_L$,
and the Poincar\'e--Cartan form $H_L$ (\ref{303})
for $L$ is the pull-back
\mar{2.32}\beq
H_L=\wh L^*H \label{2.32}
\eeq
of any such a Hamiltonian form $H$.
\end{prop}

Proposition \ref{3.22'} enables us to connect solutions of
Euler--Lagrange and Cartan 
equations for an almost regular Lagrangian $L$ with solutions of Hamilton
equations for Hamiltonian forms weakly associated with $L$.

\begin{theo}\label{3.23}  \mar{3.23}
Let a section $r$ of $\Pi\to X$
be a  solution of the Hamilton equations (\ref{b4100a}) -- (\ref{b4100b})
for a Hamiltonian form $H$ weakly associated with an almost regular
Lagrangian $L$. If $r$ lives in the Lagrangian constraint space $N_L$, the
section $s=\pi_{\Pi Y}\circ r$
of $Y\to X$ satisfies the Euler--Lagrange
equations (\ref{b327}), while $\ol s=\wh H\circ r$ obeys the Cartan equations
(\ref{b336}). 
\end{theo}

\begin{proof}
The proof is based on the relation 
$\ol L=(J^1\wh L)^*L_H$
where $\ol L$ is the Lagrangian (\ref{cmp80}) on $J^1J^1Y$ and $L_H$ is the
Lagrangian (\ref{Q3}) on $J^1\Pi$ \cite{book,jpa}.
\end{proof}

The converse assertion is more intricate \cite{book,jpa}.

\begin{theo}\label{3.24} \mar{3.24} Given an almost regular Lagrangian $L$,
let a section $\ol s$ of the jet bundle
$J^1Y\to X$ be a solution of the
Cartan equations (\ref{b336}).
Let $H$ be a Hamiltonian form weakly associated with $L$,  and let $H$ satisfy
the relation
\mar{2.36}\beq
\wh H\circ \wh L\circ \ol s=J^1(\pi^1_0\circ\ol s).\label{2.36}
\eeq
Then, the section $r=\wh L\circ \ol s$
of the Legendre bundle $\Pi\to X$ is a solution of the
Hamilton equations (\ref{b4100a}) -- (\ref{b4100b}) for $H$.
\end{theo}
 
\begin{cor} \label{ten6} \mar{ten6}
Theorems \ref{3.23}, \ref{3.24} show that, if a solution $\ol s$ of
the Cartan equations provides a solution of the covariant Hamilton equations,
its projection $\pi^1_0\circ s$ onto $Y$ is a solution of the
Euler--Lagrange equations.  
\end{cor}

Corollary \ref{ten6} gives a solution of the so-called 'second order
equation problem' in the case of  almost regular Lagrangians.

We will say that a set of Hamiltonian forms
$H$ weakly associated with an almost regular Lagrangian $L$ is
complete if, for each
solution
$s$ of the Euler-Lagrange equations, there exists a solution
$r$ of the Hamilton equations for a Hamiltonian form $H$ from this 
set such that $s=\pi_{\Pi Y}\circ r$.
By virtue of Theorem \ref{3.24}, a set
of weakly associated Hamiltonian forms
is complete if, for every solution $s$ of the Euler--Lagrange
equations for $L$, there is a Hamiltonian form $H$ from this
set which fulfills the relation 
\mar{2.36'}\beq
\wh H\circ \wh L\circ J^1s=J^1s. \label{2.36'}
\eeq

In accordance with Proposition \ref{ten4}, on an open neighbourhood in
$\Pi$ of each point $q\in N_L$, there
exists a complete set of local Hamiltonian forms weakly associated with
an almost regular Lagrangian $L$.

One  may conclude from Theorem \ref{3.24} that the
covariant Hamilton equations contain additional conditions
in comparison with the Euler--Lagrange equations. In the
case of an almost regular Lagrangian, one can introduce the constrained
Hamilton equations which are weaker than the Hamilton
equations restricted to  the Lagrangian constraint space.
Let the fibred manifold (\ref{cmp12}) admits a
global section
$\Psi$. We consider the pull-back 
\mar{b4300}\beq
H_N=\Psi^*H_L, \label{b4300}
\eeq
called the constrained Hamiltonian form. By virtue of Lemma \ref{3.22},
it does not depend on the choice of a
section $\Psi$ of the fibred manifold $J^1Y\to N_L$, and so $H_L=\wh L^*H_N$.
For sections
$r$ of the fibre bundle $N_L\to X$, we can write the 
constrained Hamilton equations
\mar{N44}\beq
r^*(u_N\rfloor dH_N) =0, \label{N44}
\eeq
where $u_N$ is an arbitrary vertical vector field on $N_L\to X$. These
equations possess the following important properties \cite{book,jpa}.

\begin{theo} \label{cmp22} \mar{cmp22} For any Hamiltonian form $H$ weakly
associated with an almost regular Lagrangian $L$, every solution $r$
of the Hamilton equations which lives in the Lagrangian constraint space $N_L$
is a solution of the constrained Hamilton equations (\ref{N44}).
\end{theo}

\begin{proof}
The proof is based on the fact that such a Hamiltonian form $H$ defines 
the global section $\wh H\circ i_N$ of the fibred manifold (\ref{cmp12}),
and $H_N=i^*_NH$.
Then the constrained Hamilton
equations can be written as
\mar{N44'}\beq
r^*(u_N\rfloor di^*_NH)=r^*(u_N\rfloor dH\mid_{N_L}) =0. \label{N44'}
\eeq
They are obviously weaker than the Hamilton equations 
(\ref{N7}) restricted to $N_L$. 
\end{proof}

\begin{theo}\label{3.02} \mar{3.02}
The constrained Hamilton equations (\ref{N44}) are equivalent to the
Hamilton--De Donder equations (\ref{N46}).
\end{theo}

\begin{proof}
By virtue of the equality (\ref{ten3}), the projection $\pi_{Z\Pi}$
(\ref{b418'}) yields a surjection of $Z_L$ onto $N_L$. Given a section 
$\Psi$ of the fibred manifold (\ref{cmp12}), we have the morphism
$\wh H_L\circ \Psi: N_L\to Z_L$.
In accordance with Lemma (\ref{3.22}), this is a surjection  such that 
\be
\pi_{Z\Pi}\circ\wh H_L\circ \Psi=\id_{N_L}.
\ee
Hence, $\wh H_L\circ \Psi$ is a bundle isomorphism over $Y$ which is
independent of the choice of a global section $\Psi$. Combining (\ref{cmp14})
and (\ref{b4300}) gives $H_N=(\wh H_L\circ \Psi)^*\Xi_L$
that leads to the desired equivalence.
\end{proof}

Since $Z_L$ and $N_L$ are isomorphic,
the Legendre morphism $H_L$ fulfills
the conditions of Theorem \ref{ddd}. Then Theorems \ref{ddd},
\ref{3.02} lead to the following assertion.

\begin{theo}\label{3.01} \mar{3.01} Let $L$ be an almost regular Lagrangian
such that the fibred manifold (\ref{cmp12}) has a global section. A section
$\ol s$ of the jet bundle
$J^1Y\to X$ is a solution of the Cartan equations (\ref{C28}) iff
$\wh L\circ
\ol s$ is a solution of  the constrained Hamilton equations (\ref{N44}).
\end{theo}

\begin{rem}
The constrained Hamiltonian form $H_N$ (\ref{b4300}) defines the
constrained Lagrangian 
\mar{cmp81}\beq
L_N=h_0(H_N)=(J^1i_N)^*L_H \label{cmp81}
\eeq
on the jet manifold $J^1N_L$ of the fibre bundle $N_L\to X$.
Then Theorem \ref{3.01} follows from the relations
\be
\ol L=(J^1\wh L)^*L_N, \qquad L_N=(J^1\Psi)^*\ol L, 
\ee
where $\ol L$ is the Lagrangian (\ref{cmp80}).
The Euler--Lagrange equation for the constrained Lagrangian $L_N$ (\ref{cmp81})
are equivalent to the constrained Hamilton equations. 
\end{rem}

\bigskip\bigskip

\begin{center}

{\large \bf PART II. Quadratic degenerate systems}

\end{center}
\bigskip\bigskip

Let us study in detail the physically important case of almost regular
quadratic Lagrangians.

\section{Lagrangian constraints}

Given a fibre bundle $Y\to X$,
let us consider a  quadratic Lagrangian $L$ which has the coordinate
 expression
\mar{N12}\beq
\cL=\frac12 a^{\la\m}_{ij} y^i_\la y^j_\m +
b^\la_i y^i_\la + c, \label{N12}
\eeq
where $a$, $b$ and $c$ are local functions on $Y$. This property is
coordinate-independent. The associated Legendre map 
\mar{N13}\beq
p^\la_i\circ\wh L= a^{\la\m}_{ij} y^j_\m +b^\la_i \label{N13}
\eeq
is an affine morphism over $Y$. It defines the corresponding linear
morphism
\mar{N13'}\beq
\ol L: T^*X\op\otimes_YVY\op\to_Y\Pi,\qquad p^\la_i\circ\ol
L=a^{\la\m}_{ij}\ol y^j_\m, \label{N13'}
\eeq
where $\ol y^j_\mu$ are  bundle coordinates on the vector
bundle $T^*X\op\ot_Y VY$.

Let the Lagrangian $L$ (\ref{N12}) be almost regular, i.e.,
the matrix function $a^{\la\m}_{ij}$ is of constant rank. Then
the Lagrangian constraint space $N_L$ 
(\ref{N13}) is an affine subbundle of the Legendre bundle $\Pi\to Y$, modelled
over the vector subbundle $\ol N_L=\im \ol L$ of  $\Pi\to
Y$. 
Hence, $N_L\to Y$ has a global section $r$. For the sake of simplicity, let us
assume that $r=0$ is the canonical
zero section $\wh 0(Y)$ of $\Pi\to Y$. Then $\ol N_L=N_L$.
Accordingly, the kernel
of the Legendre map (\ref{N13})  is an affine
subbundle of the affine jet bundle $J^1Y\to Y$, modelled over the kernel of
the linear morphism $\ol L$ (\ref{N13'}). Hence, there exists a connection 
\mar{N16,250}\ben
&&\G: Y\to \Ker\wh L\subset J^1Y, \label{N16}\\
&& a^{\la\m}_{ij}\G^j_\m + b^\la_i =0, \label{250}
\een
on $Y\to X$.
Connections (\ref{N16}) constitute an affine space modelled over the linear
space of soldering forms $\f$ on $Y\to X$ satisfying the conditions
\mar{cmp21}\beq
a^{\la\m}_{ij}\f^j_\m =0, \qquad \f^i_\la b^\la_i=0. \label{cmp21}
\eeq

\begin{rem}
In the general case of $r\neq 0$, one can consider connections 
$\G$ with values in Ker$_r\wh L$, i.e.,
\be
a^{\la\m}_{ij}\G^j_\m + b^\la_i =r^\la_i,
\ee
and replace $b$ with $b-r$ in all further constructions.
\end{rem}

The following Theorem is the key point of our analysis of quadratic
degenerate systems \cite{jpa,book00}.

\begin{theo}\label{04.2} \mar{04.2}  There exists a linear bundle
map
\mar{N17}\beq
\si: \Pi\op\to_Y T^*X\op\otimes_YVY, \qquad
\ol y^i_\la\circ\si =\si^{ij}_{\la\m}p^\m_j, \label{N17}
\eeq
such that $\ol L\circ\si\circ i_N= i_N$.
\end{theo}

The map (\ref{N17}) is a solution of the pointwise algebraic equations
\mar{N45}\beq
a\circ\si\circ a=a, \qquad 
a^{\la\mu}_{ij}\si^{jk}_{\mu\al}a^{\al\nu}_{kb}=a^{\la\nu}_{ib}.
\label{N45}
\eeq
Moreover, $\si\circ a=a\circ\si$ and 
$\si$ splits into the sum $\si=\si_0+\si_1$ of two
terms $\si_0$ and $\si_1$ satisfying the relations
\mar{N21}\beq
\si_0=\si_0\circ a\circ\si_0, \qquad a\circ\si_1=\si_1\circ a=0.
\label{N21}
\eeq
This splitting follows from the fact that the matrix $a$ in the
Lagrangian (\ref{N12}) can be seen as a global section of constant rank
of the tensor bundle 
\be
\op\w^n T^*X\op\ot_Y[\op\vee^2(TX\op\ot_Y V^*Y)]\to Y,
\ee
and there exists the bundle splitting
\be
T^*X\op\ot_Y VY=\Ker a\op\oplus_Y E.
\ee
Then $\si_0$ is a uniquely defined section of the fibre bundle 
$\op\w^n TX\op\ot_Y(\op\vee^2 E)\to Y$, while $\si_1$ is an
arbitrary sections of $\op\w^n TX\op\ot_Y(\op\vee^2 \Ker a)\to Y$.

\begin{rem}
In view of the relations (\ref{N21}), the above assumption that the
Lagrangian constraint space $N_L\to Y$ admits a global zero section
takes the form $b=(a\circ\si)b$.
\end{rem}

\begin{theo} \label{ten2} \mar{ten2}
With the relations (\ref{250}), (\ref{N45}) and (\ref{N21}), we obtain
the decompositions
\mar{N18,b4122}\bea
&& J^1Y=\cS(J^1Y)\op\oplus_Y \cF(J^1Y)=\Ker\wh L\op\oplus_Y{\rm Im}(\si\circ
\wh L), \label{N18} \\
&& y^i_\la=\cS^i_\la+\cF^i_\la= [y^i_\la
-\si^{ik}_{\la\al} (a^{\al\m}_{kj}y^j_\m + b^\al_k)]+
[\si^{ik}_{\la\al} (a^{\al\m}_{kj}y^j_\m + b^\al_k)], \label{b4122}
\eea
\mar{N20,'}\bea
&& \Pi=\cR(\Pi)\op\oplus_Y\cP(\Pi)=\Ker\si_0 \op\oplus_Y N_L, \label{N20} \\
&& p^\la_i = \cR^\la_i+\cP^\la_i= [p^\la_i -
a^{\la\m}_{ij}\si^{jk}_{\m\al}p^\al_k] +
[a^{\la\m}_{ij}\si^{jk}_{\m\al}p^\al_k]. \label{N20'}
\eea
\end{theo}

With respect to the coordinates $\cS^i_\la$,
$\cF^i_\la$ (\ref{b4122}) and $\cR^\la_i$, $\cP^\la_i$ (\ref{N20'}),
the Lagrangian (\ref{N12}) reads  
\mar{cmp31}\beq
\cL=\frac12 a^{\la\m}_{ij}\cF^i_\la\cF^j_\m +c', \label{cmp31}
\eeq
while the Lagrangian constraint space is given by the reducible constraints
\mar{zzz}\beq
\cR^\la_i= p^\la_i -
a^{\la\m}_{ij}\si^{jk}_{\m\al}p^\al_k=0. \label{zzz}
\eeq

Note that, in gauge theory, we have the canonical splitting (\ref{N18}) where
$2\cF$ is the strength tensor \cite{book}.  The
Yang--Mills Lagrangian of gauge theory is exactly of the form (\ref{cmp31})
where $c'=0$. The Lagrangian of Proca fields is also of the form (\ref{cmp31})
where
$c'$ is the mass term. This is an example of a  degenerate
Lagrangian system without gauge symmetries.

Given the linear map $\si$ (\ref{N17}) and a connection $\G$
(\ref{N16}), let us consider the affine Hamiltonian map
\mar{N19}\beq
\Phi_{\si\G}=\G\circ\pi_{\Pi Y}+\si:\Pi \op\to J^1Y,  \qquad
\Phi_{\si\G}{}^i_\la = \G^i_\la  + \si^{ij}_{\la\m}p^\m_j. \label{N19}
\eeq
It satisfies the relation (\ref{2.30a}). Then 
the Hamiltonian form
\mar{N22}\ben
&& H_{\si\G}=H_{\Phi_{\si\G}} +\Phi^*_{\si\G}L= p^\la_idy^i\w\om_\la -
[\G^i_\la p^\la_i +\frac12 \si_0{}^{ij}_{\la\m}p^\la_ip^\m_j
+\si_1{}^{ij}_{\la\m}p^\la_ip^\m_j -c']\om=
\label{N22}\\
&& \quad (\cR^\la_i+\cP^\la_i)dy^i\w\om_\la - [(\cR^\la_i+\cP^\la_i)\G^i_\la
+\frac12
\si_0{}^{ij}_{\la\m}\cP^\la_i\cP^\m_j
+\si_1{}^{ij}_{\la\m}\cR^\la_i\cR^\m_j -c']\om,\nonumber
\een
is weakly associated with the Lagrangian $L$
(\ref{N12}) in accordance with Proposition \ref{jp}. The corresponding
Lagrangian (\ref{Q3}) reads
\mar{ten7}\beq
L_H=[(\cR^\la_i+\cP^\la_i)(y^i_\la-\G^i_\la)
-\frac12
\si_0{}^{ij}_{\la\m}\cP^\la_i\cP^\m_j
-\si_1{}^{ij}_{\la\m}\cR^\la_i\cR^\m_j +c']\om. \label{ten7}
\eeq

\begin{theo} \label{cmp30}  \mar{cmp30} 
Given a linear map $\si$ (\ref{N17}), the Hamiltonian forms $H_{\si\G}$
(\ref{N22}) 
parametrized by connections $\G$ (\ref{N16}) constitute a complete set. 
\end{theo}

\begin{proof}
Let us consider the Hamilton equations
(\ref{b4100a}), written as the equality
\mar{N10}\beq
J^1(\pi_{\Pi Y}\circ r)= \wh H\circ r \label{N10}
\eeq
for
a section $r$ of the Legendre bundle $\Pi\to X$.
The Hamiltonian map $\wh H_{\si\G}$ reads
\be
\wh H_{\si\G}=\Phi_{\si\G}+\frac12\si_1=\G\circ\pi_{\Pi Y}+\si+\si_1.
\ee
Due to the projections $\cS$, $\cF$ (\ref{b4122}),
the Hamilton equations (\ref{N10}) break in two parts
\mar{N23}\ben
&&\cS\circ J^1(\pi_{\Pi Y}\circ r)=\G, \label{N23}\\
&& (\dl - \si a)^{\m i}_{j\la}(\dr_\m r^j-\G^j_\m)=0, \nonumber
\een
\mar{N28}\ben
&&\cF \circ J^1(\pi_{\Pi Y}\circ r)=\si+\si_1, \label{N28}\\
&& (\si a)^{\m i}_{j\la}(\dr_\m r^j-\G^j_\m)=0. \nonumber
\een
Let $s$ be an arbitrary section of $Y\to X$,
e.g., a solution of the Euler--Lagrange
equations. There exists a connection $\G$ (\ref{N16}) such
that the relation (\ref{N23}) holds, namely, $\G={\cal S}\circ\G'$ where
$\G'$ is a
connection on $Y\to X$ which has $s$ as an integral section. 
It is easily seen that, in this case, the Hamiltonian map (\ref{N19})
satisfies the relation (\ref{2.36'})
for $s$.
Hence, the Hamiltonian forms (\ref{N22}) constitute
a complete set.
\end{proof}

 We have different complete sets of Hamiltonian
forms (\ref{N22}) for different $\si_1$. For instance, if $\si_1=0$, then
$\Phi_{\si\G}=\wh H_{\si\G}$ and the 
Hamiltonian forms (\ref{N22}) are associated with the Lagrangian (\ref{N12}).
If $\si_1$ is non-degenerate, so are the Hamiltonian forms (\ref{N22}).
Hamiltonian forms $H$
(\ref{N22}) of a complete set in Theorem \ref{cmp30} differ from each
other in the term 
$\f^i_\la\cR^\la_i$, where $\f$ are the soldering forms (\ref{cmp21}). 
This term vanishes on the
Lagrangian constraint space (\ref{zzz}). Accordingly, the constrained
Hamiltonian form reads
\be
H_N=i_N^*H_{\si\G}=\cP^\la_idy^i\w\om_\la - [\cP^\la_i\G^i_\la
+\frac12
\si_0{}^{ij}_{\la\m}\cP^\la_i\cP^\m_j-c'],
\ee
and the  constrained Hamilton equations (\ref{N44}) can be
written. In the case of quadratic Lagrangians, we can improve Theorem
\ref{cmp22} as follows \cite{book,jpa}.

\begin{theo} \label{cmp23} \mar{cmp23} For every Hamiltonian
form $H_{\si\G}$ (\ref{N22}),
the Hamilton equations (\ref{b4100b}) and (\ref{N28}) restricted to the
Lagrangian constraint space $N_L$  are equivalent to the constrained Hamilton
equations.
\end{theo}

It follows that, restricted to the Lagrangian constraint
space, the Hamilton equations for different Hamiltonian forms (\ref{N22})
associated with the same quadratic Lagrangian (\ref{N12}) differ from each
other in the equations (\ref{N23}). These equations are independent of 
momenta and play the role of gauge-type conditions.

Note that, in Hamiltonian gauge theory, the restricted Hamiltonian form and
the restricted Hamilton equations are gauge invariant, while weakly
associated Hamiltonian forms (\ref{N22}) and Lagrangians (\ref{ten7}) contain
gauge fixing terms. Moreover, one can find a complete set of non-degenerate
Hamiltonian forms, that is
essential for quantization.

\section{Geometry of antighosts}

Using the splitting (\ref{N20}) and the corresponding projection
operators
\mar{yy,xxx}\ben
&& P^{\la k}_{i \nu}= a_{ij}^{\la\m}\si_0{}^{jk}_{\m\nu},
\qquad  R^{\la k}_{i \nu}=(\dl_i^k\dl^\la_\nu -
a_{ij}^{\la\m}\si_0{}^{jk}_{\m\nu}), \label{yy}\\
&& P^{\la k}_{i \nu}\cR^\nu_k=0, \qquad R^{\la k}_{i \nu}\cR^\nu_k=\cR^\la_i,
\label{xxx}
\een
we can construct the Koszul--Tate
resolution for the Lagrangian constraints (\ref{zzz}) of a generic almost
regular quadratic Lagrangian (\ref{N12}) in an explicit form.
Since these constraints are reducible, one needs
an infinite number of antighost fields in general \cite{fisch,kimura} (we
follow the terminology of Ref. \cite{kimura}). They are graded by the antighost
number
$r$ and the Grassmann parity $r\,{\rm mod}2$. Odd antighost fields are
represented by elements of a simple graded manifold \cite{book00}. To
describe even antighost fields, we should generalize
the notion of a graded manifold  to commutative graded
algebras generated  both by odd and even elements \cite{mpl,jmp00,book00}.

Let $E=E_0\oplus E_1\to Z$ be the Whitney sum of vector bundles $E_0\to
Z$ and $E_1\to Z$ over a manifold $Z$. One can think of $E$ as
being a bundle of vector superspaces with a typical fibre $V=V_0\oplus
V_1$. Let us consider 
the exterior bundle
\be
\w E^*=\op\bigoplus^\infty_{k=0} (\op\w^k_Z E^*)
\ee
which is the tensor bundle $\ot E^*$ modulo  elements
\be
e_0e'_0 - e'_0e_0, \quad e_1e'_1 + e'_1e_1, \quad e_0e_1 - e_1e_0\quad
e_0,e'_0\in E_{0z}^*,
\quad e_1,e'_1\in E_{1z}^*, \quad z\in Z.
\ee
Global sections of $\w E^*$ constitute a graded commutative algebra
$\cA(Z)$ which is the product of the commutative algebra $\cA_0(Z)$ of
global sections of 
the symmetric bundle $\vee E_0^*\to Z$ and the graded algebra $\cA_1(Z)$ of
global sections of the exterior bundle $\w E_1^*\to Z$.
The pair
$(Z,\cA_1(Z))$ is a (simple) graded manifold \cite{bart,book00}.  For
the sake of brevity, we 
agree to call
$(Z,\cA(Z))$ a graded commutative manifold. Accordingly, elements of $A(Z)$ are called graded
commutative functions. 
Let $\{c^a\}$ be the holonomic bases for $E^*\to Z$ with respect to some bundle
atlas $(z^A,v^i)$ of $E\to Z$ with transition functions $\{\rho^a_b\}$, i.e.,
$c'^a=\rho^a_b(z)c^b$. Then graded commutative functions read
\mar{z785}\beq
f=\op\sum_{k=0} \frac1{k!}f_{a_1\ldots
a_k}c^{a_1}\cdots c^{a_k}, \label{z785}
\eeq
where $f_{a_1\cdots
a_k}$ are local functions on $Z$, and we omit the symbol of an exterior product
of elements $c$.

Let us introduce the differential calculus in these functions.
We start from the $\cA(Z)$-module $\der \cA(Z)$ of graded derivations of
the graded commutative algebra $\cA(Z)$. They are defined as
endomorphisms of $\cA(Z)$ such that 
\mar{mm81}\beq
 u(ff')=u(f)f'+(-1)^{\nw u\nw f}fu (f') \label{mm81}
\eeq
for homogeneous elements $u\in \der\cA(Z)$ and $f,f'\in \cA(Z)$.
We use the notation $\nw.$ for the Grassmann parity.
Due to the canonical splitting
$VE= E\times E$, the vertical tangent bundle
$VE\to E$ can be provided with the fibre bases $\{\dr_a\}$ dual of $\{c^a\}$.
These are fibre bases for $\pr_2VE=E$. Then
any derivation $u$ of $\cA(U)$ on a trivialization domain $U$ of $E$ reads
\mar{mm83}\beq
u= u^A\dr_A + u^a\dr_a, \label{mm83}
\eeq
where $u^A, u^a$ are local graded commutative functions and $u$ acts on
$f\in \cA(U)$ by the rule
\be
u(f_{a_1\cdots
a_k}c^{a_1}\cdots c^{a_k})=u^A\dr_A(f_{a_1\cdots
a_k})c^{a_1}\cdots c^{a_k} +u^a
f_{a_1\ldots a_k}\dr_a\rfloor (c^{a_1}\cdots c^{a_k}).
\ee
This rule implies the corresponding
coordinate transformation law
\mar{lmp2}\beq
u'^A =u^A, \qquad u'^a=\rho^a_ju^j +u^A\dr_A(\rho^a_j)c^j \label{lmp2}
\eeq
of derivations (\ref{mm83}).
Let us consider
the vector bundle
$\cV_E\to Z$ which is locally isomorphic to the vector bundle
\be
\cV_E\mid_U\approx\w E^*\op\ot_Z(\pr_2VE\op\oplus_Z TZ)\mid_U,
\ee
and has the transition functions
\be
&& z'^A_{i_1\ldots i_k}=\rho^{-1}{}_{i_1}^{a_1}\cdots
\rho^{-1}{}_{i_k}^{a_k} z^A_{a_1\ldots a_k}, \\
&& v'^i_{j_1\ldots j_k}=\rho^{-1}{}_{j_1}^{b_1}\cdots
\rho^{-1}{}_{j_k}^{b_k}\left[\rho^i_jv^j_{b_1\ldots b_k}+ \frac{k!}{(k-1)!}
z^A_{b_1\ldots b_{k-1}}\dr_A(\rho^i_{b_k})\right]
\ee
of the bundle coordinates $(z^A_{a_1\ldots a_k},v^i_{b_1\ldots b_k})$,
$k=0,\ldots$. It is readily observed that, for any
trivialization domain $U$, the
$\cA$-module $\der\cA(U)$ with the transition functions (\ref{lmp2}) is
isomorphic to the $\cA$-module of local sections of $\cV_E\mid_U\to U$.
One can show that, if $U'\subset U$ are open
sets, there is the restriction morphism $\der\cA(U)\to
\der\cA(U')$. It follows that, restricted to an open subset $U$, every
derivation $u$ of
$\cA(Z)$ coincides with some local section $u_U$ of $\cV_E\mid_U\to U$, whose
collection $\{u_U, U\subset Z\}$ defines uniquely a global section of
$\cV_E\to Z$, called a graded vector field on $Z$.

The $\w E^*$-dual $\cV^*_E$ of $\cV_E$ is a vector bundle over $Z$
whose sections
constitute the $\cA(Z)$-module of graded
1-forms $\f=\f_A dz^A + \f_adc^a$.
Then
the morphism $\f:u\to \cA(Z)$ can be seen as the interior product
\beq
u\rfloor \f=u^A\f_A + (-1)^{\nw{\f_a}}u^a\f_a. \label{cmp65}
\eeq
Graded $k$-forms $\f$ are defined as sections
of the graded exterior bundle $\w^k_Z\cV^*_E$ such that
\be
 \f\w\si =(-1)^{\nm\f\nm\si +\nw\f\nw\si}\si \w \f,
\ee
where $|.|$ is the form degree.
The interior product (\ref{cmp65})
is extended to higher graded forms by the rule
\be
u\rfloor (\f\w\si)=(u\rfloor \f)\w \si
+(-1)^{\nm\f+\nw\f\nw{u}}\f\w(u\rfloor\si).
\ee
The graded exterior differential
$d$ of graded functions is introduced by the condition
$u\rfloor df=u(f)$
for an arbitrary graded vector field $u$, and  is
extended uniquely to higher graded forms by the rules
\be
d(\f\w\si)= (d\f)\w\si +(-1)^{\nm\f}\f\w(d\si), \qquad  d\circ d=0.
\ee

\section{The Koszul--Tate resolution}

Let us turn to the splitting (\ref{N20}) and the projection operators
(\ref{yy}).
To construct the vector bundle $E$ of antighosts, let
us consider the vertical tangent bundle $V_Y\Pi$ of $\Pi\to Y$.
Let us chose the bundle $E$ as the Whitney sum of the
bundles $E_0\oplus E_1$ over $\Pi$ which are the infinite Whitney sum
over
$\Pi$ of the copies of
$V_Y\Pi$.  We have
\be
E= V_Y\Pi\op\oplus_\Pi V_Y\Pi\op\oplus_\Pi\cdots.
\ee
This bundle is provided with the holonomic coordinates $(x^\la,y^i,p_i^\la,\dot
p_i^{\la(r)})$, $r=0,1,\ldots$, where  $(x^\la,y^i,p_i^\la,\dot
p_i^{\la(2l)})$ are coordinates on $E_0$, while
$(x^\la,y^i,p_i^\la,\dot p_i^{\la(2l+1)})$ are those on $E_1$. By
$r$ is meant the antighost number.  The
dual of $E\to \Pi$ is
\be
E^*= V^*_Y\Pi\op\oplus_\Pi V^*_Y\Pi\op\oplus_\Pi\cdots.
\ee
It is  endowed with the
associated fibre bases
$\{c_i^{\la(r)}\}$, $r=1,2,\ldots$, such that
$c_i^{\la(r)}$ have the same linear coordinate
transformation law as the coordinates $p_i^\la$. The corresponding
graded vector fields and graded forms are introduced on $\Pi$ as sections of
the vector bundles $\cV_E$ and $\cV^*_E$, respectively.

The $C^\infty(\Pi)$-module $\cA(\Pi)$ of graded functions is graded by the
antighost number as
\be
\cA(\Pi)=\op\oplus_{r=0}^\infty \cN^r, \qquad \cN^0=C^\infty(\Pi).
\ee
Its terms $\cN^r$ constitute a complex
\mar{mm90}\beq
0\leftarrow C^\infty(\Pi) \leftarrow \cN^1 \leftarrow \cdots \label{mm90}
\eeq
with respect to the Koszul--Tate  differential
\mar{mm91}\ben
&& \dl: C^\infty(V^*Y)\to 0, \nonumber \\
&& \dl(c^{\la(2l)}_i)=P^{\la k}_{i \nu}c^{\nu(2l-1)}_k,
\qquad l>0,
\label{mm91}\\
&& \dl(c^{\la(2l+1)}_i)=R^{\la k}_{i \nu} c^{\nu(2l)}_k,
\qquad l>0,
\nonumber\\ 
&&  \dl(c^{\la(1)}_i)=R^{\la k}_{i \nu} p_k^\nu.
\nonumber
\een
The nilpotency property $\dl\circ\dl=0$ of this differential is the corollary
of the relations (\ref{xxx}).

It is readily observed that the complex (\ref{mm90}) with respect to the
differential (\ref{mm91}) has the homology
groups
\be
H_{k>0}=0, \qquad H_0=C^\infty(\Pi)/I_{N_L}=C^\infty(N_L),
\ee
where $I_{N_L}$ is an ideal of smooth functions on $\Pi$ which vanish on
the Lagrangian constraint space $N_L$. Thus, this is a desired Koszul--Tate
resolution of the Lagrangian constraints (\ref{zzz}).

Note that, in different particular cases of the degenerate quadratic Lagrangian
(\ref{N12}), the complex (\ref{mm90}) may have a subcomplex, which is also
the Koszul--Tate resolution. For instance, if the matrix $a$ is diagonal with
respect to some adapted coordinates on $J^1Y$, the constraints
(\ref{zzz}) are irreducible and the complex (\ref{mm90}) contains a
subcomplex which consists only of the antighosts
$c_i^{\la(1)}$.

\bigskip\bigskip

\begin{center}

{\large \bf PART III. Constraints in time-dependent mechanics}

\end{center}
\bigskip\bigskip

If $X=\bR$, polysymplectic Hamiltonian formalism provides the adequate
Hamiltonian formulation of time-dependent mechanics \cite{book98,sard98}.
Here, we study holonomic time-dependent constraints \cite{jmp00}.

Note that, in contrast with the existent
formulations of time-dependent mechanics, we do not imply any
preliminary splitting of its momentum phase space  
$\Pi=\bR\times Z$. From the physical viewpoint, this splitting characterizes
a certain reference frame, and is violated by  time-dependent transformations.
Given such a splitting, $\Pi$ is
endowed with the product of the zero Poisson structure on
$\bR$ and the Poisson structure on $Z$. A
Hamiltonian $\cH$ is defined as a real function  on
$\Pi$. The corresponding Hamiltonian vector field $\vt_\cH$ on 
$\Pi$ is vertical with respect to
the fibration $\Pi\to\bR$. Due to the natural imbedding 
$\Pi\times_\bR T\bR\to T\Pi$
one introduces the vector field $\g_\cH=\dr_t +\vt_\cH$,
where $\dr_t$ is the standard vector field on $\bR$. The Hamilton
equations are equations for the integral curves of the vector field
$\g_H$, while the evolution equation on the Poisson algebra
$C^\infty(\Pi)$ of smooth functions on $\Pi$ is given by the Lie derivative
\be
\bL_{\g_\cH}f= \dr_t f +\{\cH,f\}. 
\ee
However, the splitting in the
right-hand side of this expression is violated by
time-dependent transformations, and a Hamiltonian
$\cH$ is not scalar under these transformations. Its Poisson bracket with
functions $f\in C^\infty(\Pi)$ is ill-defined, and is not maintained under
time-dependent transformations. This fact is the key point of the study of
constraints in Hamiltonian time-dependent mechanics.

\section{Hamiltonian time-dependent mechanics}

Let us consider time-dependent mechanics on
a configuration bundle $Q\to\bR$. 

\begin{rem}
The following peculiarities of fibre bundles over $\bR$ should be
emphasized \cite{book98}. Their base $\bR$ is parametrized by the
Cartesian coordinates
$t$ with the transition functions 
$t'=t+$const., and  is
provided with the standard vector field $\dr_t$ and the
standard 1-form
$dt$. A vector field $u$ on a fibre bundle $Y\to\bR$ is said to be projectable
if
$u\rfloor dt$ is constant. From now on, by vector fields on fibre bundles
over $\bR$ are meant only projectable vector fields.
Let
$Y\to\bR$ be a fibre bundle coordinated by $(t,y^A)$ and
$J^1Y$ its first order jet manifold, equipped with the adapted coordinates
$(t,y^A,y^A_t)$. There is the canonical imbedding 
$J^1Y\to TY$ over $Y$ 
whose image is the affine subbundle 
of elements $\up\in TY$ such that
$\up\rfloor dt=1$. This subbundle is modelled over the vertical tangent bundle
$VY\to Y$. As a consequence, there is one-to-one correspondence between the
connections on the fibre bundle $Y\to \bR$ and the 
vector fields $\G$ on $Y$ such that $\G\rfloor dt=1$. 
A
connection $\G$ on $Y\to\bR$ 
yields a 1-dimensional distribution on $Y$,
transversal to the fibration $Y\to\bR$. As a consequence,
it defines an atlas of local constant
trivializations of
$Y\to\bR$ whose
transition functions are independent of $t$ and  
$\G=\dr_t$. Conversely, 
every atlas of local constant trivializations of a fibre  bundle
$Y\to\bR$ sets a connection on  $Y\to\bR$ which is $\dr_t$
relative to this atlas.
In particular, 
every trivialization of $Y\to \bR$ yields a complete
connection $\G$ on $Y$, and {\it vice versa}. 
\end{rem}

The momentum 
phase space of time-dependent mechanics is the vertical cotangent bundle
\be
V^*Q\ar^{\pi_Q} Q\ar^\pi\bR
\ee 
endowed with holonomic coordinates $(t,q^i,p_i)$. The homogeneous
Legendre bundle $Z_Q$ (\ref{N41}) is the cotangent bundle $T^*Q$.
The $V^*Q$ is provided with the canonical Poisson structure $\{,\}_V$ such
that 
\mar{m72'}\beq
\zeta^*\{f,g\}_V=\{\zeta^*f,\zeta^*g\}_T, \qquad f,g\in
C^\infty(V^*Q),\label{m72'}
\eeq
where $\zeta=\pi_{Z\Pi}$ (\ref{b418'}) is the natural fibration
\mar{mm5}\beq
\zeta:T^*Q\to V^*Q, \label{mm5}
\eeq
and $\{,\}_T$ is the canonical Poisson structure on the cotangent
bundle $T^*Q$ provided with the symplectic form $d\Xi$.
The characteristic
distribution of $\{,\}_V$ coincides with the
vertical tangent bundle $VV^*Q$ of $V^*Q\to \bR$. 

Given  a section $h$ of the  fibre bundle (\ref{mm5}), let us consider the
pull-back forms
\mar{z401}\beq
\bth =h^*(\Xi\w dt), \qquad \bom=h^*(d\Xi\w dt) \label{z401}
\eeq
on $V^*Q$. It is readily observed that these forms are independent of $h$, and
are canonical on $V^*Q$. 
Then a Hamiltonian vector field  
$\vt_f$ for a function $f$ on $V^*Q$ is given by the relation
\be
\vt_f\rfloor\bom = -df\w dt,
\ee
while the Poisson bracket (\ref{m72'}) is written as
\be
\{f,g\}_Vdt=\vt_g\rfloor\vt_f\rfloor\bom.
\ee 
Thus, the 3-form $\bom$ (\ref{z401}) provides $V^*Q$ with the Poisson structure
$\{,\}_V$ in an equivalent way. 
Furthermore,  holonomic coordinates on $V^*Q$
are canonical for the Poisson structure (\ref{m72'}) such that 
\mar{m72}\ben
&& \bom = dp_i\w dq^i\w dt, \nonumber \\
&&\{f,g\}_V=\dr^if\dr_ig-\dr^ig\dr_if, \qquad f,g\in C^\infty(V^*Q).
\label{m72}
\een 

\begin{lem} \label{jmp1} \mar{jmp1} \cite{book98,sard98}.
A vector field $u$ on $V^*Q$ is canonical for the Poisson structure $\{,\}_V$
iff the form $u\rfloor\bom$ is closed.  The closed form
$u\rfloor\bom$ is exact.
\end{lem}

With respect to the Poisson bracket (\ref{m72}), the Hamiltonian vector field  
$\vt_f$ for a function $f$ on the Legendre bundle $V^*Q$ is 
\be
\vt_f = \dr^if\dr_i- \dr_if\dr^i. 
\ee
It is vertical. Conversely, one can show that every vertical canonical
vector field on the Legendre bundle
$V^*Q\to\bR$ is locally a Hamiltonian vector field.

\begin{prop}\label{ex3} \mar{ex3}
Let a connection $\g$  on the
Legendre bundle $V^*Q\to\bR$ be a canonical vector field for the Poisson
structure $\{,\}_V$.  Then 
$\g\rfloor\bom=dH$ where $H$ is locally a Hamiltonian form. 
  Conversely, any Hamiltonian form 
\mar{b4210}\beq
H=h^*\Xi= p_i dq^i -\cH dt \label{b4210}
\eeq
on the momentum phase space
$V^*Q$ admits a unique Hamiltonian connection 
\mar{m57}\beq
\g_H=\dr_t +\dr^i\cH\dr_i -\dr_i\cH\dr^i. \label{m57}
\eeq
\end{prop}

\begin{rem}
A glance at the expression (\ref{b4210}) shows that, given a trivialization
of the configuration bundle $Q\to\bR$, the Hamiltonian form
$H$ (\ref{b4210}) is the well-known integral
invariant of Poincar\'e--Cartan where $\cH$ plays the role of a Hamiltonian.
\end{rem}

Hamiltonian forms in time-dependent mechanics constitute an affine space
modelled over the vector space of horizontal densities $fdt$ on $V^*Q\to\bR$,
i.e., over $C^\infty(V^*Q)$. Accordingly, Hamiltonian
connections $\g_H$ make up an affine space modelled over the vector 
space of Hamiltonian vector fields. 
Every Hamiltonian form yields 
the  Hamiltonian map 
\mar{mm41}\beq
\wh H=J^1\pi_Q\circ\g_H:V^*Q\to J^1Q, \qquad q^i_t\circ\wh H= \dr^i\cH.
\label{mm41}
\eeq

In particular, let $\G$ be a connection on $Q\to\bR$. It  characterizes a
reference frame in time-dependent mechanics
\cite{eche95,book98,massa,sard98}, and defines the frame Hamiltonian form
\be
H_\G=p_idq^i -p_i\G^idt. 
\ee
The corresponding Hamiltonian connection is the canonical lift
\be
V^*\G=\dr_t +\G^i\dr_i -p_i\dr_j\G^i \dr^j
\ee
of $\G$ onto $V^*Q\to \bR$.
Then  any Hamiltonian form $H$ on $V^*Q$ admits the splittings
\mar{m46'}\beq
H=H_\G -\wt\cH_\G dt, \qquad \cH= p_i\G^i + \wt\cH_\G, \label{m46'}
\eeq
where 
$\wt\cH_\G$ is the energy
function with respect to the reference frame $\G$ \cite{book98,sard98}.

Given a Hamiltonian form $H$ (\ref{b4210}) and the associated
Hamiltonian connection
$\g_H$ (\ref{m57}), the kernel of the covariant differential
$D_{\g_H}$ defines the Hamilton
equations 
\mar{m41}\beq
q^i_t =\dr^i\cH, \qquad p_{ti} =-\dr_i\cH. \label{m41}
\eeq

A Hamiltonian form $H$
(\ref{b4210}) is the Poincar\'e--Cartan form for the Lagrangian 
\be
L_H=h_0(H) = (p_iq^i_t - \cH)dt 
\ee
on the jet manifold $J^1V^*Q$. This Lagrangian is a convenient tool
in order to apply the standard Lagrangian technique to Hamiltonian
time-dependent mechanics. 
As in the polysymplectic case, the
Hamilton equations (\ref{m41}) for $H$
are exactly the Lagrange equations
for $L_H$. 
Furthermore, given a function $f\in C^\infty(V^*Q)$ and its
pull-back onto $J^1V^*Q$, let us consider the bracket
\be
(f,L_H)=\dl^i f \dl_i L_H -
\dl_i f \dl^i L_H = \bL_{\g_H}f- d_t f,
\ee
where $\dl^i$, $\dl_i$
are variational derivatives
(in the spirit of the Batalin--Vilkovisky antibracket). Then
the equation 
$(f,L_H)=0$ is the evolution equation  
\mar{jmp4}\beq
d_tf=\bL_{\g_H}f= \dr_tf +\{\cH,f\}_V \label{jmp4}
\eeq
in time-dependent mechanics. Note that, taken separately, the terms in its
right-hand side are ill-behaved  objects under reference
frame  transformations. 
With the splitting (\ref{m46'}), the evolution equation (\ref{jmp4}) is
brought into the frame-covariant form
\be
\bL_{\g_H}f= V^*\G\rfloor H +\{\wt\cH_\G,f\}_V, 
\ee
but its right-hand side does not reduce to a Poisson bracket.  

The following construction enables us to represent
the right-hand side of the evolution equation (\ref{jmp4}) as a pure Poisson
bracket. Given a Hamiltonian form $H=h^*\Xi$, let us consider its pull-back
$\zeta^*H$ onto the cotangent bundle $T^*Q$. It is readily observed that the
difference
$\Xi-\z^*H$ is a horizontal 1-form on $T^*Q\to\bR$, while  
\mar{mm16}\beq
\cH^*=\dr_t\rfloor(\Xi-\zeta^*H)=p+\cH \label{mm16}
\eeq
is a function on $T^*Q$. Then the relation
\mar{mm17}\beq
\z^*(\bL_{\g_H}f)=\{\cH^*,\z^*f\}_T \label{mm17}
\eeq
holds for any function $f\in C^\infty(V^*Q)$. In particular, $f$ is an
integral of motion iff its bracket (\ref{mm17}) vanishes.
Note that $\g_H=T\z(\vt_{\cH^*})$ where $\vt_{\cH^*}$
be the Hamiltonian vector field for the function $\cH^*$ (\ref{mm16}) with
respect to the canonical Poisson structure $\{,\}_T$ on $T^*Q$.

\section{Time-dependent constraints} 

With the Poisson bracket $\{,\}_V$, an algebra of time-dependent constraints 
can be described similarly to that in
conservative Hamiltonian mechanics, but we should use the relation
(\ref{mm17}) in order to extend the constraint algorithm to time-dependent
constraints. 

Let $N$ be a closed imbedded subbundle $i_N:N\op\hookrightarrow V^*Q$ of the
Legendre bundle
$V^*Q\to\bR$, treated as a constraint space. 
Note that $N$ is
neither Lagrangian nor symplectic submanifold with respect to the Poisson
structure $\{,\}_V$.
Let us consider the ideal
$I_N\subset C^\infty(V^*Q)$ of functions
$f$ on
$V^*Q$ which vanish on $N$, i.e.,
$i_N^*f=0$. Its elements are said to be constraints. There is the isomorphism 
\mar{gm93}\beq
C^\infty(V^*Q)/I_N\cong C^\infty(N)\label{gm93}
\eeq 
of associative commutative algebras. 
By the normalize $\ol I_N$ of the ideal
$I_N$ is meant the subset of functions of $C^\infty(V^*Q)$ whose Hamiltonian
vector fields restrict to vector fields on $N$ \cite{kimura}, i.e., 
\mar{gm95}\beq
\ol I_N=\{f\in C^\infty(V^*Q):\,\, \{f,g\}_V\in I_N, \,\, \forall g\in I_N\}.
\label{gm95}
\eeq
It follows from the Jacobi identity that the normalizer
(\ref{gm95}) is a Poisson subalgebra of
$C^\infty(V^*Q)$. Put
\mar{gm96}\beq
I'_N= \ol I_N\cap I_N. \label{gm96}
\eeq
This is also a Poisson subalgebra of $\ol I_N$. Its elements are called 
the first class constraints, while
the remaining elements of $I_N$ are the second class constraints. It is
readily observed that $I^2_N\subset I'_N$.

\begin{rem}
Let $N$ be a coisotropic
submanifold of $V^*Q$. Then $I_N\subset
\ol I_N$ and $I_N=I'_N$, i.e., all constraints are of first class.
\end{rem}

Let $H$ be a Hamiltonian form on the momentum phase space $V^*Q$.
In accordance with the relation (\ref{mm17}), a constraint $f\in I_N$ is
preserved with respect to a Hamiltonian form $H$ if the bracket (\ref{mm17})
vanishes on the constraint space. It follows that solutions of the Hamilton
equations (\ref{m41}) do not leave the constraint space $N$
if 
\mar{mm20}\beq
\{\cH^*,\z^*I_N\}_T\subset \z^*I_N. \label{mm20}
\eeq
If this relation does not  hold, let us introduce  
secondary constraints $\{\cH^*,\zeta^*f\}_T$, $f\in I_N$, which belong to
$\zeta^* (C^\infty(V^*Q))$. If the set of primary and secondary
constraints is not closed  with respect to the relation (\ref{mm20}), one can
add the tertiary constraints $\{\cH^*,\{\cH^*,\zeta^*f_a\}_T\}_T$, and so
on.

Let us assume that $N$ is a final constraint space for a Hamiltonian form
$H$.  If $H$ satisfies the relation (\ref{mm20}), so is a
Hamiltonian form
\mar{mm21}\beq
H_f=H-fdt \label{mm21}
\eeq
where $f\in I'_N$ is a first class constraint. 
Though Hamiltonian forms $H$ and $H_f$ coincide
with each other on the constraint space $N$, the corresponding Hamilton
equations have different solutions in $N$ because
$dH\mid_N\neq dH_f\mid_N$. At the same time, $d(i_N^*H)=d(i_N^*H_f)$.
Therefore, let us consider the constrained Hamiltonian
form
\mar{mm23}\beq
H_N=i_N^*H_f \label{mm23}
\eeq
which is the same for all $f\in I'_N$.
Note that $H_N$ (\ref{mm23}) is not a true Hamiltonian form on
$N\to \bR$ in general.
On sections $r$ of the bundle $N\to\bR$, we can write the
constrained Hamilton equations
\mar{NN44}\beq
r^*(u_N\rfloor dH_N) =0, \label{NN44}
\eeq
where $u_N$ is an arbitrary vertical vector field on $N\to \bR$. 
It is readily observed that,
for any Hamiltonian form $H_f$
(\ref{mm21}), every solution
of the Hamilton equations which lives in the constraint space
$N$ is a solution of the constrained Hamilton equations (\ref{NN44}).

Let us mention the problem  of constructing a generalized Hamiltonian
system, similar to that for a Dirac constraint system in conservative
mechanics. Let $H$ satisfy the condition
$\{\cH^*,\zeta^*I'_N\}_T\subset I_N$, whereas
$\{\cH^*,\zeta^*I_N\}_T\not\subset I_N$. The goal is to find 
a constraint $f\in I_N$  such that the modified Hamiltonian $H -fdt$ would
satisfy the condition
\be
\{\cH^* +\zeta^*f ,\zeta^* I_N\}_T\subset \zeta^*I_N.
\ee
This is an equation for a second-class constraint
$f$. 

The above construction, except the isomorphism
(\ref{gm93}), can be applied to any ideal $I$ of $C^\infty(V^*Q)$,
treated as an ideal of constraints \cite{kimura}. In particular, an
ideal $I$ is said 
to be coisotropic if it is a Poisson algebra. In this case, $I$ is a Poisson
subalgebra of the normalize
$\ol I$ (\ref{gm95}), and coincides with $I'$ (\ref{gm96}).

Note that, since $\zeta^*(\bL_{\vt_f}H)\neq
\{\zeta^*f,\cH^*\}_T$, the constraints $f\in I_N$ preserved with respect to a
Hamiltonian form $H$ (i.e. $\{\zeta^*f,\cH^*\}_T\in I_N$) are not generators
of gauge symmetries of $H$ in general. 
At the same time, the generators
of gauge symmetries of a Hamiltonian form $H$ define an ideal of constraints as
follows. Let $\cA$ be a Lie algebra of generators $u$ of gauge symmetries of a
Hamiltonian form $H$. The
corresponding symmetry currents 
$J_u=u\rfloor H$
on
$V^*Q$ constitute a Lie algebra with
respect to the Poisson bracket 
\be
\{J_u,J_{u'}\}=J_{[u,u']}
\ee
on $V^*Q$. Let $I_\cA$ denotes the
ideal of $C^\infty(V^*Q)$ generated by these symmetry currents. 
It is readily observed that this ideal is coisotropic. Then one can think of
$I_\cA$ as being an ideal of first class constraints compatible with the
Hamiltonian form $H$, i.e., 
\beq
\{\cH^*,\zeta^*I_\cA\}_T\subset \zeta^*I_\cA. \label{mm33}
\eeq
Note that any Hamiltonian form $H_u=H-J_udt$,
$u\in\cA$, obeys the same relation (\ref{mm33}), but other currents $J_{u'}$
are not conserved with respect to $H_u$, unless $[u,u']= 0$.

\section{BRST charge for Lagrangian constraints}

Lagrangian constraints in time-dependent mechanics are described in the
same manner as in the general polysymplectic case \cite{jmp00}. At the
same time, the canonical Poisson structure on the momentum phase space
$V^*Q$ enables us to construct the BRST charge for the Koszul-Tate
differential.

In time-dependent mechanics, the vector bundle $E$ of antighosts for
Lagrangian constraints is the infinite Whitney sum
\be
E= V_Q(V^*Q)\op\oplus_{V^*Q}V_Q(V^*Q)\oplus\cdots
\ee
over
$V^*Q$ of the copies of
$V_Q(V^*Q)$.  
This bundle is provided with the holonomic coordinates $(t,q^i,p_i,\dot
p_i^{(r)})$, $r=0,1,\ldots$, where  $(t,q^i,p_i,\dot
p_i^{(2l)})$ are coordinates on $E_0$, while 
$(t,q^i,p_i,\dot p_i^{(2l+1)})$ are those on $E_1$. The
dual of
$E\to V^*Q$ is
\be
E^*= V^*_Q(V^*Q)\op\oplus_{V^*Q}V^*_Q(V^*Q)\oplus\cdots.
\ee
It is  endowed with the
associated fibre bases
$\{c_i^{(r)}\}$, $r=1,2,\ldots$, such that
$c_i^{(r)}$ have the same linear coordinate
transformation law as the coordinates $p_i$. The corresponding
graded vector fields and graded forms are introduced on $V^*Q$ as sections of
the vector bundles $\cV_E$ and $\cV^*_E$, respectively. 
The $C^\infty(V^*Q)$-module $\cA(V^*Q)$ of graded commutative functions 
is graded by the antighost number $r$.
Its terms $\cN^r$ constitute the Koszul--Tate resolution (\ref{mm90})
with respect to the Koszul--Tate  differential
\be
&& \dl: C^\infty(V^*Q)\to 0, \nonumber \\
&& \dl(c^{(2l)}_i)= a_{ij}\si^{jk}_0c^{(2l-1)}_k, \qquad l>0, \\
&& \dl(c^{(2l+1)}_i)=(\dl_i^k- a_{ij}\si^{jk}_0)c^{(2l)}_k, \qquad, l>0,
\nonumber\\ 
&&  \dl(c^{(1)}_i)=(\dl_i^k- a_{ij}\si^{jk}_0)p_k. 
\ee

Let us construct the BRST
charge $\bQ$ such that 
\be
\dl(f)=\{\bQ, f\}, \qquad f\in \cA(V^*Q),
\ee
with respect to some Poisson bracket. The problem is to find the Poisson
bracket such that $\{f,g\}=0$ for all $f,g\in C^\infty(V^*Q)$.

To overcome
this difficulty, one can consider the vertical extension of Hamiltonian
formalism onto the configuration bundle $VQ\to\bR$ \cite{jmp99,book98,jmp00}.
The corresponding
Legendre bundle $V^*(VQ)$  is isomorphic to
$V(V^*Q)$, and is provided with the holonomic coordinates $(t,q^i,p_i,\dot
q^i,\dot p_i)$ such that $(q^i,\dot p_i)$ and $(\dot q^i, p_i)$ are 
conjugate pairs of canonical coordinates. The momentum phase space $V(V^*Q)$ is
endowed with the canonical exterior 3-form
\beq
\bom_V=\dr_V\bom=[d\dot p_i\w dq^i +dp_i\w d\dot q^i]\w dt,
\label{m145}
\eeq
where we use the compact notation
\be
\dot\dr_i=\frac{\dr}{\dr\dot q^i}, \quad \dot\dr^i=\frac{\dr}{\dr\dot
p_i}, \quad \dr_V=\dot q^i\dr_i +\dot p_i\dr^i.
\ee
The corresponding Poisson bracket on $V(V^*Q)$ reads
\be
\{f,g\}_{VV} =\dot\dr^if\dr_ig +\dr^if\dot\dr_ig -\dr^ig\dot\dr_if
-\dot\dr^ig\dr_if. 
\ee

To extend this bracket to graded functions, let us consider the following
graded extension of Hamiltonian formalism \cite{gozz,book98,book00,ijmp}.
We will assume that $Q\to\bR$ is a vector bundle, and will further denote
$\Pi=V^*Q$.

Let us consider the vertical tangent bundle $VV\Pi$. It
admits the canonical decomposition 
\mar{cmp68}\beq
VV\Pi=V\Pi\op\oplus_\bR V\Pi\ar^{\pr_1} V\Pi. \label{cmp68}
\eeq
Let choose the bundle $E$ as the
Whitney sum of the bundles $E_0\oplus E_1$ over $V\Pi$ which are the
infinite Whitney sum over
$V\Pi$ of the copies of
$VV\Pi$. In view of the decomposition (\ref{cmp68}), we have
\be
E=V\Pi\op\oplus_\Pi V\Pi\oplus\cdots\op\to^{\pr_1} V\Pi.
\ee
This bundle is provided with the holonomic coordinates $(t,q^i,p_i,\dot
q^i_{(r)},\dot p_i^{(r)})$, $r=0,1,\ldots$, where  $(t,q^i,p_i,\dot
q^i_{(2l)},\dot p_i^{(2l)})$ are coordinates on $E_0$ and 
$(t,q^i,p_i,\dot
q^i_{(2l+1)},\dot p_i^{(2l+1)})$ are those on $E_1$.  The dual
of
$E\to V\Pi$ is
\be
E^*=V\Pi\op\oplus_\bR V\Pi^*
\oplus\cdots.
\ee
It is  endowed with the
associated fibre bases
$\{\ol c^i_{(r)},\ol c_i^{(r)},c^i_{(r)},c_i^{(r)}\}$, $r=1,\ldots$. The
corresponding graded vector fields and graded forms are introduced on $V\Pi$ as
sections of the vector bundles $\cV_E$ and $\cV^*_E$, respectively. Let us
complexify these bundles as $\bC\op\ot_\bR\cV_{VV\Pi}$ and
$\bC\op\ot_\bR\cV^*_{VV\Pi}$.  

The BRST extension of the form (\ref{m145}) on $V^*Q$ is the 3-form 
\be
\bom_S=\bom_V +i\op\sum_{r=1}^\infty(d\ol
c_i^{(r)}\w dc^i_{(r)}- dc_i^{(r)}\w d\ol c^i_{(r)})\w dt 
\ee
The corresponding bracket of graded functions on $V^*Q$ reads
\mar{mm103}\ben
&&\{f,g\}_S=\{f,g\}_{VV} -i\op\sum_{r=1}^\infty (-1)^{r[f]}[
\frac{\dr f}{\dr \ol c_i^{(r)}}\frac{\dr g}{\dr c^i_{(r)}} + (-1)^r
\frac{\dr f}{\dr \ol c^i_{(r)}}\frac{\dr g}{\dr c_i^{(r)}}- \label{mm103}\\
&&\qquad \frac{\dr f}{\dr  c_i^{(r)}}\frac{\dr g}{\dr \ol c^i_{(r)}} - (-1)^r
\frac{\dr f}{\dr  c^i_{(r)}}\frac{\dr g}{\dr \ol c_i^{(r)}}]. \nonumber
\een
It satisfies the condition $\{f,g\}_S=-(-1)^{[f][g]}\{g,f\}_S$.
Then the desired  BRST charge takes the form
\be
\bQ=i[\ol c^i_{(1)}(\dl_i^k - a_{ij}\si^{jk}_0)p_k + \op\sum_{l=1}^\infty (\ol
c^i_{(2l)}a_{ij}\si^{jk}_0c_k^{(2l-1)} + \ol
c^i_{(2l+1)}(\dl_i^k-a_{ij}\si^{jk}_0)c_k^{(2l)})].
\ee
Due to the bracket (\ref{mm103}), one can use this charge in order to obtain
the BRST complex for antighosts
$c_i^{(r)}$ and ghosts $\ol c^i_{(r)}$ such that
\be
\ol c^i_{(2l-1)}\mapsto a_{kj}\si^{ij}_0\ol c^k_{(2l)}, \qquad 
\ol c^i_{(2l)}\mapsto -(\dl^i_k-a_{rj}\si^{ij}_0)\ol c^k_{(2l+1)}, \qquad l>0.
\ee

\end{document}